\documentclass[conference]{IEEEtran}
\usepackage{cite}

\ifCLASSINFOpdf
\else
   \usepackage[dvips]{graphicx}
   \DeclareGraphicsExtensions{.eps}
\fi
\usepackage[cmex10]{amsmath}
\usepackage{array}

\usepackage[tight,footnotesize]{subfigure}

\usepackage{amssymb}

\newtheorem{thm}{Theorem}
\newtheorem{lem}[thm]{Lemma}

\newtheorem{ex}{Example}

\begin{document}
%
\title{Evolutionary Approaches To Minimizing\\Network Coding Resources}

\author{
Minkyu Kim, Muriel M\'{e}dard, Varun Aggarwal, Una-May O'Reilly, Wonsik Kim,\\Chang Wook Ahn, and Michelle Effros
\thanks{M. Kim, M. M\'{e}dard, and W. Kim are with the Laboratory of Information and Decision Systems, Massachusetts Institute of Technology, Cambridge, MA 02139, USA (\{minkyu, medard, wskim14\}@mit.edu).}
\thanks{V. Aggarwal and U.-M. O'Reilly are with the Computer Science and Artificial Intelligence Laboratory, Massachusetts Institute of Technology, Cambridge, MA 02139, USA (\{varun\_ag@, unamay@csail.\}mit.edu).}
\thanks{C. W. Ahn is with the Department of Information and Communications, Gwangju Institute of Science and Technology, Gwangju 500-712, Korea (cwan@evolution.re.kr).}
\thanks{M. Effros is with the Data Compression Laboratory, California Institute of Technology, Pasadena, CA 91125, USA (effros@caltech.edu).}
}

\IEEEoverridecommandlockouts


\maketitle

\begin{abstract}
We wish to minimize the resources used for network coding while achieving the desired throughput in a multicast scenario. We employ evolutionary approaches, based on a genetic algorithm, that avoid the computational complexity that makes the problem NP-hard. Our experiments show great improvements over the sub-optimal solutions of prior methods. Our new algorithms improve over our previously proposed algorithm in three ways. First, whereas the previous algorithm can be applied only to acyclic networks, our new method works also with networks with cycles. Second, we enrich the set of components used in the genetic algorithm, which improves the performance. Third, we develop a novel distributed framework. Combining distributed random network coding with our distributed optimization yields a network coding protocol where the resources used for coding are optimized in the setup phase by running our evolutionary algorithm at each node of the network. We demonstrate the effectiveness of our approach by carrying out simulations on a number of different sets of network topologies.
\end{abstract}


%
\IEEEpeerreviewmaketitle

\section{Introduction} \label{sec:Intro}

It is now well known that network throughput can be significantly increased by employing the novel technique of network coding, where the intermediate nodes are allowed to combine data received from different links \cite{ACLY00, LYC03}. While most network coding solutions employ coding at all possible nodes, it is often possible to achieve the network coding advantage by coding only at a subset of nodes.

\begin{figure}[h]
\centerline{
\subfigure[Network $B$]{\includegraphics[height=1.2in]{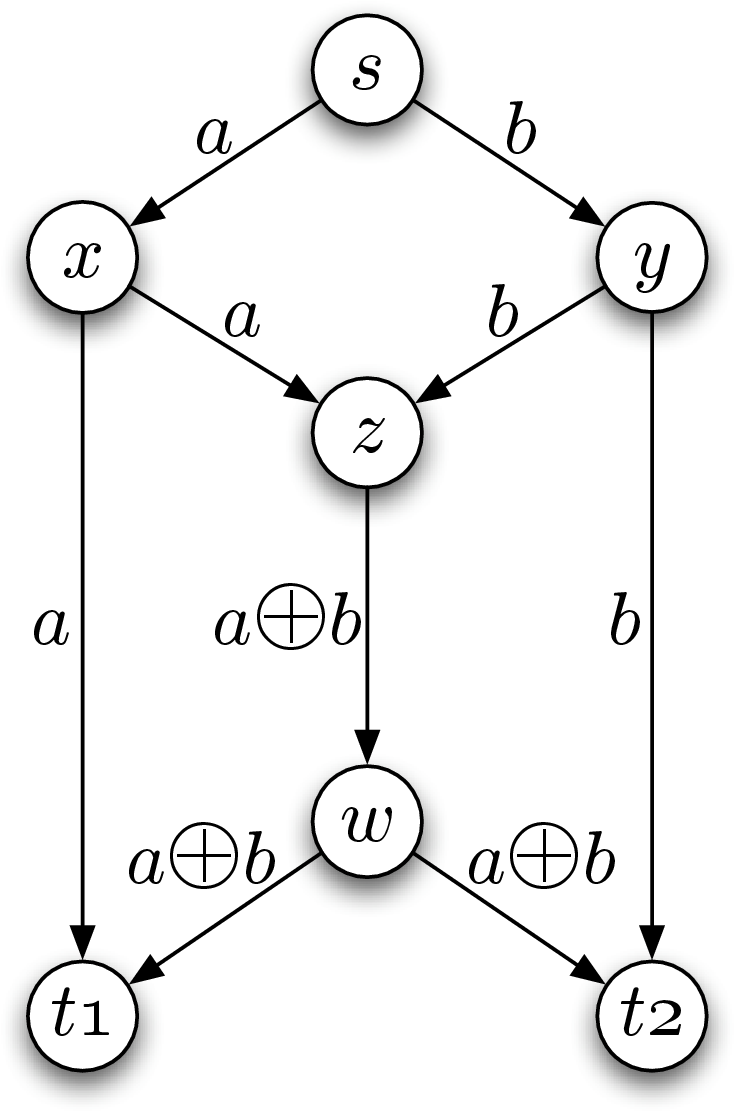}
\label{fig:bf}}
\hfil
\subfigure[Network $B'$]{\includegraphics[height=1.2in]{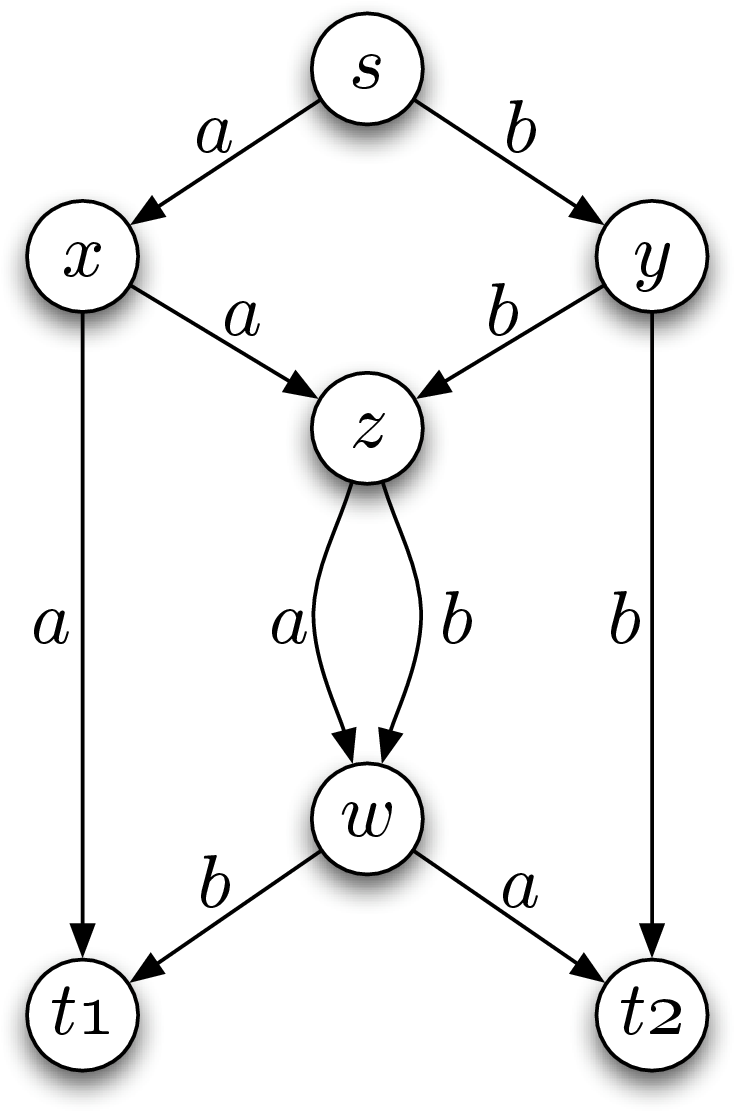}
\label{fig:bf2}}
\hfil
\subfigure[Network $C$]{\includegraphics[height=1.2in]{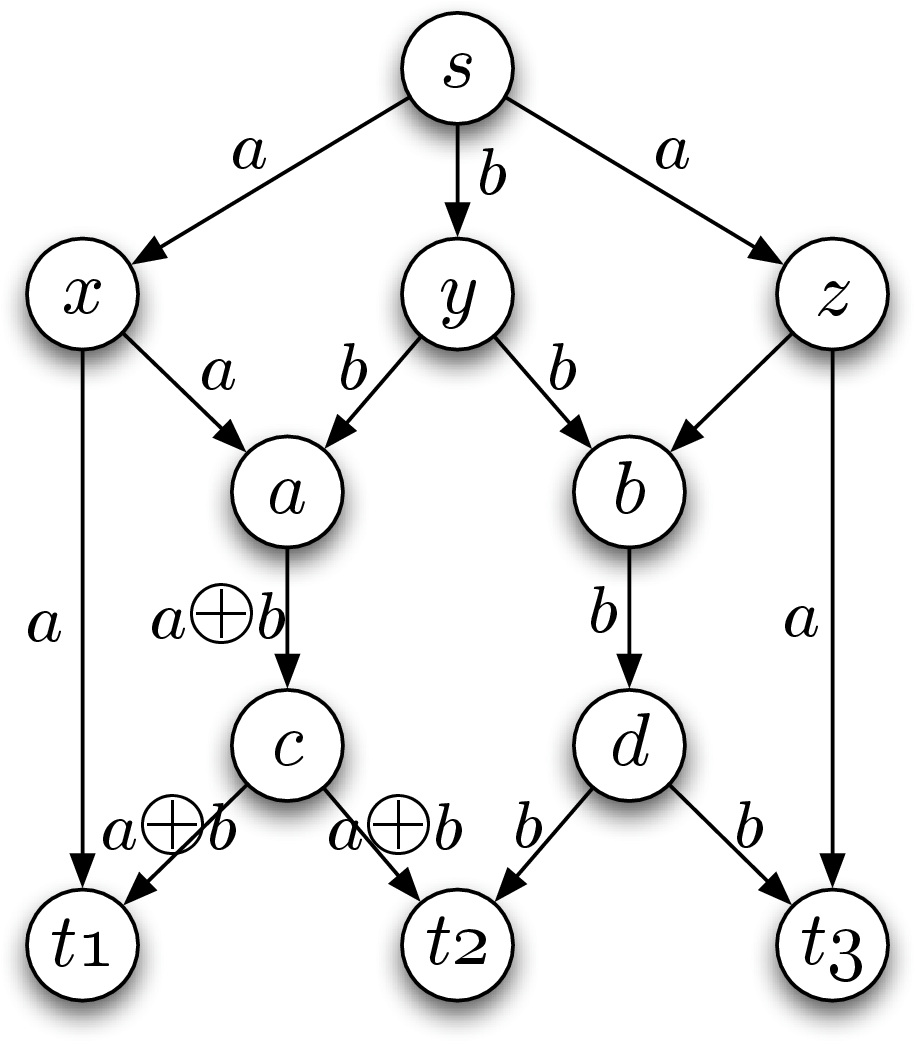}
\label{fig:corr}}
}
\caption{Sample Networks for Example \ref{ex:intro}}
\label{fig:sample}
\vspace{-0.1in}
\end{figure}

\begin{ex}
In the canonical example of network $B$ (Fig. \ref{fig:bf}) \cite{ACLY00}, only node $z$ needs to combine its two inputs while all other nodes perform routing only. If we suppose that link $(z,w)$ in network $B$ has capacity 2, which we represent by two parallel unit-capacity links in network $B'$ (Fig. \ref{fig:bf2}), a multicast of rate 2 is possible without network coding. In network $C$ (Fig. \ref{fig:corr}), where node $s$ wishes to transmit data at rate 2 to the 3 leaf nodes, network coding is required at either node $a$ or node $b$, but not both. \hfill $\square$
\label{ex:intro}
\end{ex}

Example \ref{ex:intro} leads us to the following question: At which nodes does network coding need to occur to achieve the multicast capacity? If network coding is handled at the application layer, we can minimize the cost of network coding by identifying the nodes where access up to the application layer is not necessary. If network coding is integrated in the buffer management of a router, it is important to understand where and how many such special routers must be deployed to satisfy the communication demands.

Determining a minimal set of nodes where coding is required is difficult. The problem of deciding whether a given multicast rate is achievable without coding, i.e., whether the minimum number of required coding nodes is zero or not, reduces to a multiple Steiner subgraph problem, which is NP-hard \cite{RP86}. Hence, the optimization problem to find the minimal number of required coding nodes is NP-hard. Even approximating the minimal number of coding nodes within any multiplicative factor or within an additive factor of $|V|^{1-\epsilon}$ is NP-hard \cite{LSB06}.

In \cite{KAME06}, we introduce an evolutionary approach for finding a practical multicast protocol that provides the full benefit of network coding with a reduced number of coding nodes. The proposed approach uses a Genetic Algorithm (GA) that operates on a set of candidate solutions which it improves sequentially via mechanisms inspired by biological evolution (e.g., recombination/mutation of genes and survival of the fittest). The algorithm proposed in \cite{KAME06} reduces the number of coding links/nodes relative to prior approaches and applies to a variety of generalized scenarios.

The new algorithm proposed here improves on the one in \cite{KAME06} in three key ways. First, whereas the algorithm in \cite{KAME06} can be applied to only acyclic networks, we devise a modified method that works also with networks with cycles. Second, we introduce a new set of GA components that in our experiments significantly outperforms the one used in \cite{KAME06}. Third, we develop a novel framework where most time consuming computations of the evolutionary algorithm are distributed over the network. This new framework, combined with the distributed random network coding scheme of \cite{HKM03}, can make a distributed network coding protocol where the resources used for coding are optimized in the setup phase as our proposed algorithm running at each node of the network.

The rest of the paper is organized as follows. Section \ref{sec:ProbRelatedWork} presents the problem formulation and summarizes related work. Section \ref{sec:Centralized} describes the improvements of the algorithm using a centralized framework. Section \ref{sec:Distributed} extends this approach to a distributed framework. Section \ref{sec:Exp} presents experimental results. Section \ref{sec:Con} concludes with topics for future research.

\section{Problem Formulation and Related Work}\label{sec:ProbRelatedWork}

\subsection{Problem Formulation}\label{sec:Prob}

We assume that the network is given by a directed multigraph $G=(V,E)$, where each link has a unit capacity. Connections with larger capacities are represented by multiple links. Only integer flows are allowed, hence there is either no flow or a unit rate of flow on each link. We consider the single multicast scenario in which a single source $s \in V$ wishes to transmit data at rate $R$ to a set $T \subset V$ of sink nodes, where $|T|=d$. Rate $R$ is said to be achievable if there exists a transmission scheme that enables all $d$ sinks to receive all of the information sent. We consider only linear coding, where a node's output on an outgoing link is a linear combination of the inputs from its incoming links. Linear coding is sufficient for multicast \cite{LYC03}.

Given the target rate $R$, which we assume is achievable if coding is allowed at all nodes, we wish to determine a minimal set of nodes where coding is required in order to achieve this rate. Coding is necessary at a node $v \in V$ if coding is necessary on at least one of node $v$'s outgoing links. As pointed out also in \cite{LSB06}, the number of coding links is a more accurate estimator of the amount of computation incurred by coding. We assume hereafter that our objective is to minimize the number of coding \emph{links} rather than \emph{nodes}. Note, however, that as demonstrated in \cite{KAME06}, it is straightforward to generalize the proposed algorithm to the case of minimizing the number of coding nodes. Furthermore, \cite{KAME06} shows that, with appropriate changes, the algorithm can be readily applied to more generalized optimization scenarios, e.g., where different links/nodes have different costs for coding.

It is clear that no coding is required at a node with only a single input since it has nothing to combine with. We refer to a node with multiple incoming links as a \emph{merging node}. If the linearly coded output on a particular outgoing link of a particular merging node weights all but one incoming message by zero, then no coding occurs on that link. (Even if the only nonzero coefficient is not identity, there is another coding scheme that replaces the coefficient by identity \cite{LSB06}.) Thus, to determine whether coding is necessary on an outgoing link of a merging node, we need to verify whether we can constrain the output on the link to depend on a single input without destroying the achievability of the given rate.

\begin{figure}[h]
\vspace{-0.1in}
\centerline{
\subfigure[Merging node $v$]{\includegraphics[height=1.4in]{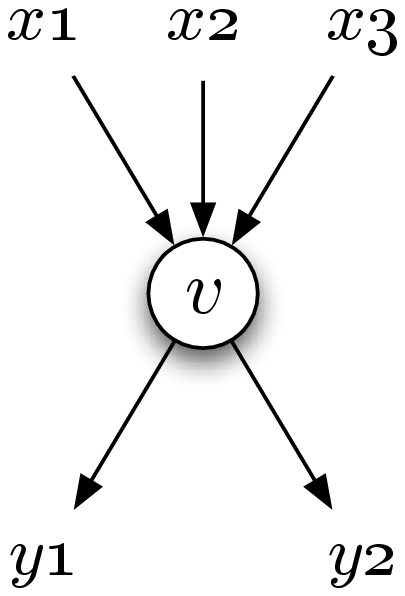}
\label{fig:block1}}
\hfil
\subfigure[Two blocks for outgoing links]{\includegraphics[height=1.4in]{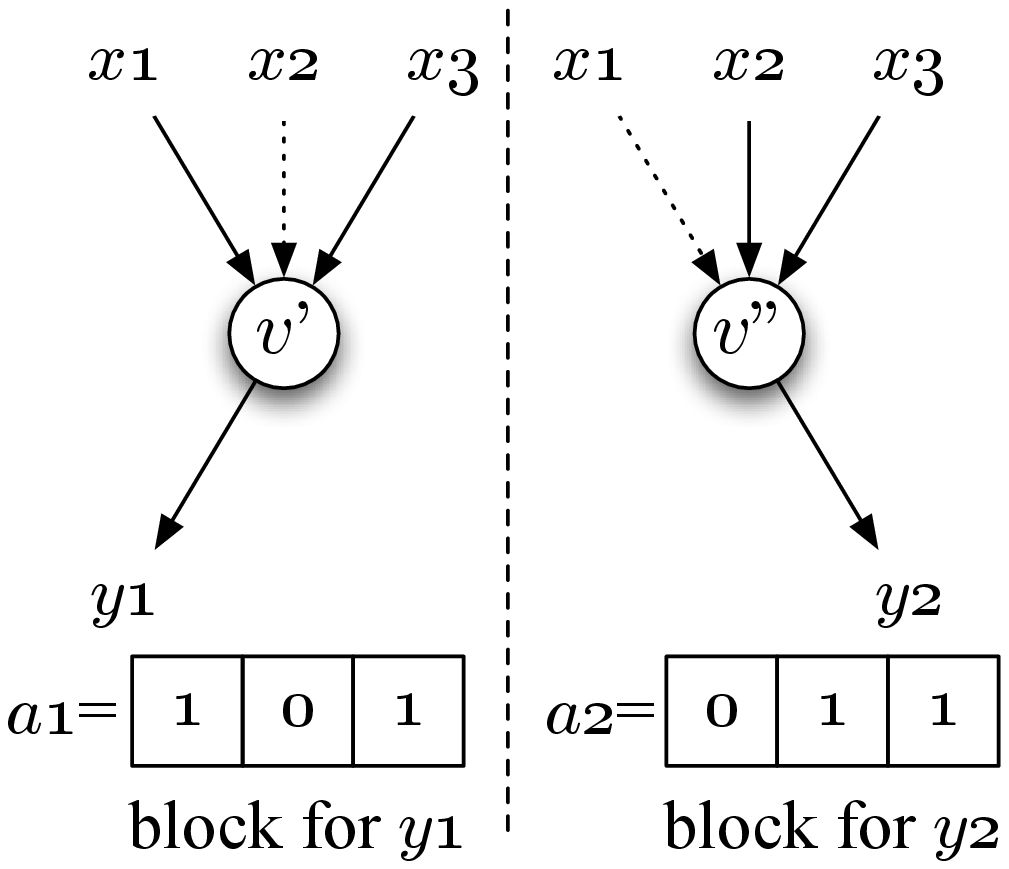}
\label{fig:block2}}
}
\caption{Node $v$ with 3 incoming and 2 outgoing links has inputs described by vectors $a_1=(a_{11}, a_{21}, a_{31})$ and $a_2=(a_{12}, a_{22}, a_{32})$.}
\label{fig:block}
\end{figure}

Consider a merging node with $k (\geq 2)$ incoming links and $l (\geq 1)$ outgoing links. For each $i \in \{1, ..., k\}$ and each $j \in \{1, ..., l\}$, we set $a_{ij}=1$ if the input from incoming link $i$ contributes to the linearly coded output on outgoing link $j$, and $a_{ij}=0$ otherwise; we call these the \emph{active} and \emph{inactive} states, respectively. Network coding is required over link $j$ only if two or more link states are active. Thus, it is useful to think of $a_j=(a_{ij})_{i\in\{1, ..., k\}}$ as a \emph{block} of length $k$ (see Fig. \ref{fig:block} for an example).

As in network $C$ of Example \ref{ex:intro}, whether node $v$ must code over link $y_j$ varies depending on which other nodes are coding. Thus deciding which nodes should code in general involves a selection out of exponentially many possible choices. We employ a GA-based search method to efficiently address the large and exponentially scaling size of the space.

\subsection{Related Work}

Fragouli et al. \cite{FS06} show that coding is required at no more than $(d-1)$ nodes in acyclic networks with 2 unit-rate sources and $d$ sinks. This result, however, is not easily generalized to more than 2 sources. They also present an algorithm to construct a minimal subtree graph. For target rate $R$, they first select a subgraph consisting of $R$ link-disjoint paths to each of $d$ sinks and then construct the corresponding labeled line graph in which they sequentially remove the links whose removal does not affect the achievable rate.

Langberg et al. \cite{LSB06} derive an upper bound on the number of required coding nodes for both acyclic and cyclic networks. They give an algorithm to construct a network code that achieves the bounds, where the network is first transformed such that each node has degree at most 3 and each of the links is sequentially examined and removed if the target rate is still achievable without it.

Both of the above approaches remove links sequentially in a greedy fashion, assuming that network coding is done at all nodes with multiple incoming links in the remaining graph. Note that, unless a good order of the link traversal is found, the quality of the solution cannot be much improved as illustrated in \cite{KAME06}.

Bhattad et al. \cite{BRKN05} give linear programming formulations for the problems of optimizing over various resources used for network coding, based on a model allowing continuous flows. Their optimal formulations, however, involve a number of variables and constraints that grows exponentially with the number of sinks, which makes it hard to apply the formulations to the case of a large number of sinks, even at the price of sacrificed optimality.

We conclude this section with a brief introduction to GA.

\subsection{A Brief Introduction to GA}\label{sec:GAIntro}

GAs \cite{Mit96} operate on a set of candidate solutions, called a \emph{population}. Each solution is typically represented by a bit string, called a \emph{chromosome}. Each chromosome is assigned a \emph{fitness value} that measures how well the chromosome solves the problem at hand, compared with other chromosomes in the population. From the current population, a new population is generated typically using three genetic operators: \emph{selection}, \emph{crossover} and \emph{mutation}. Chromosomes for the new population are selected randomly (with replacement) in such a way that fitter chromosomes are selected with higher probability. For crossover, survived chromosomes are randomly paired, and then two chromosomes in each pair exchange a subset of their bit strings to create two offspring. Chromosomes are then subject to mutation, which refers to random flips of the bits applied individually to each of the new chromosomes. The process of evaluation, selection, crossover and mutation forms one \emph{generation} in the execution of a GA. The above process is iterated with the newly generated population successively replacing the current one. The GA terminates when a certain stopping criterion is reached, e.g., after a predefined number of generations. GAs have been applied to a large number of scientific and engineering problems, including many combinatorial optimization problems in networks (e.g., \cite{ES96, DAS97}).

There are several aspects of our problem suggesting that a GA-based method may be a promising candidate: GA has proven to work well if the space to be searched is large, but known not to be perfectly smooth or unimodal, or even if the space is not well understood \cite{Mit96} (which makes traditional optimization methods difficult to apply). Note that the search space of our problem is apparently not smooth or unimodal with respect to the number of coding links and the structure of the space consisting of the feasible binary vectors is not well understood. Since the problem is NP-hard, it is not critical that the calculated solution may not be a global optimum. Note also that, while it is hard to characterize the structure of the search space, once provided with a solution we can verify its feasibility and count the number of coding links therein in polynomial time. Thus, if the use of genetic operations can suitably limit the size of the space to be actually searched, a solution can be obtained fairly efficiently.

\section{Centralized Approach}\label{sec:Centralized}

\begin{figure}[h]
    \centering
    \includegraphics[height=1.6in]{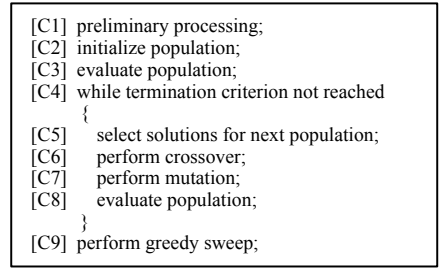}
    \vspace{-0.1in} 
    \caption{Flow of Centralized Algorithm}
    \label{fig:flowcentral}
\vspace{-0.1in}    
\end{figure}

We first present the centralized version of the algorithm, whose overall structure, based on simple GA \cite{Mit96}, is shown in Fig. \ref{fig:flowcentral}. Sections \ref{sec:Algebraic} and \ref{sec:GraphDecomp} present two different methods for mapping the network coding problem to a GA framework (procedure [C1] in Fig. \ref{fig:flowcentral}) and evaluating the chromosomes ([C3, C8] in Fig. \ref{fig:flowcentral}). Either of the two methods can be combined with the computational part of the algorithm (remaining procedures in Fig. \ref{fig:flowcentral}) which is described in Section \ref{sec:CompPartGA}.

\subsection{Algebraic Method}
\label{sec:Algebraic}

We first describe the algebraic method by which a choice of coding links is mapped to a GA problem and a given candidate solution (chromosome) is evaluated. Owing to space limitations, here we present only the main concepts; the reader is referred to \cite{KAME06} for details. This algebraic method will also be used later in the distributed version of the algorithm. This method applies only to acyclic networks; for cyclic networks, it can be very inefficient as discussed in Section \ref{sec:GraphDecomp}. 

Given an acyclic graph $G=(V,E)$, we first construct the corresponding labeled line graph $G'=(V',E')$ \cite{KM03}, where each node in $V'$ represents a link in $E$ and each link $(v',w') \in E'$ implies that the links $e, f \in E$ corresponding to nodes $v', w' \in V'$, respectively, are connected in $G$ via some node $u \in V$ such that $u=\text{head}(e)=\text{tail}(f)$. To construct a network code, we assign a coefficient to each link in $G'$ as in \cite{KM03}. Note that there is a one-to-one correspondence between the binary variables $a_{ij}$ introduced in Section \ref{sec:Prob} and the coefficients assigned to the incoming links to a node with \emph{multiple} incoming links in $G'$. Thus, for each binary variable $a_{ij}$ we can consider the associated coefficient. If there are $m$ such coefficients in $G'$, a chromosome is represented by a vector consisting of $m$ binary variables; if we denote by $d^v_{in}$ and $d^v_{out}$ the in-degree and the out-degree of node $v \in V$, $m$ is given by $m=\sum_{v \in \mathcal{V}} d^v_{in} d^v_{out}$, where $\mathcal{V} \subset V$ is the set of all merging nodes in $G$.

To evaluate a given chromosome, we first verify its feasibility. If $a_{ij}=0$ in the chromosome, then input $x_i$ is inactive with respect to output $y_j$ and we set associated coefficient to zero. If $a_{ij}=1$, then we let the associated coefficient be an indeterminate nonzero value. To determine whether the target rate $R$ is achievable, we rely on random linear coding; i.e., to each of the remaining coefficients we assign a random element from a finite field and check whether the system matrix is nonsingular. Note that this feasibility test entails a bounded error, which is shown in \cite{KAME06} not to be critical since the error is one-sided, i.e., a feasible chromosome may mistakenly be declared infeasible but not vice versa, and we can lower the bound on the error probability as much as we desire at an additional cost of computation.

We then define the fitness value $F$ of chromosome $\underline{z}$ as
\begin{equation}
F(\underline{z})=
    \begin{cases}
        \text{number of coding links}, & \text{if } \underline{z} \text{ is feasible}, \\
        \infty, & \text{if } \underline{z} \text{ is infeasible},
    \end{cases}
\label{eq:fitnessDef}
\end{equation}
where the number of coding links can be easily calculated by counting the number of blocks in the chromosome with at least two $1$'s. It is not hard to verify that the computational complexity required to evaluate a single chromosome is $O(d \cdot (|E|^{2.376}+R^3))$.

\subsection{Graph Decomposition Method}\label{sec:GraphDecomp}

Note that the above algebraic method deals explicitly with the scalar coefficients that appear in the system matrix assuming that the network operates with zero delay (and thus the network is cycle-free). In the presence of cycles, delay must be taken into account, hence the system matrix becomes a matrix over the polynomial ring with coefficients that are \emph{rational functions} in the delay variable $D$ \cite{KM03}. In this case, the matrix computation involves calculating the coefficient for each power of the delay variable $D$, which in general renders the feasibility test prohibitively inefficient.

In this subsection, we show that, with an appropriate graph decomposition, the evaluation can be done by calculating the max-flows between the source and the sinks. Note that the minimum of those max-flows equals the maximum achievable multicast rate regardless of the existence of cycles in the network \cite{ACLY00}. Unlike the algebraic one, this modified method operates on the actual graph $G$ rather than the labeled line graph.

\begin{figure}[h]
\centerline{
\subfigure[Before decomposition]{\includegraphics[height=1in]{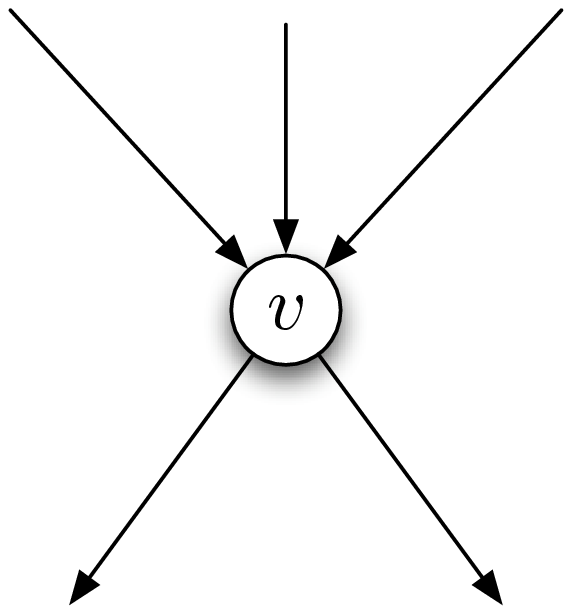}
\label{fig:decomp1}}
\hfil
\subfigure[After decomposition]{\includegraphics[height=1in]{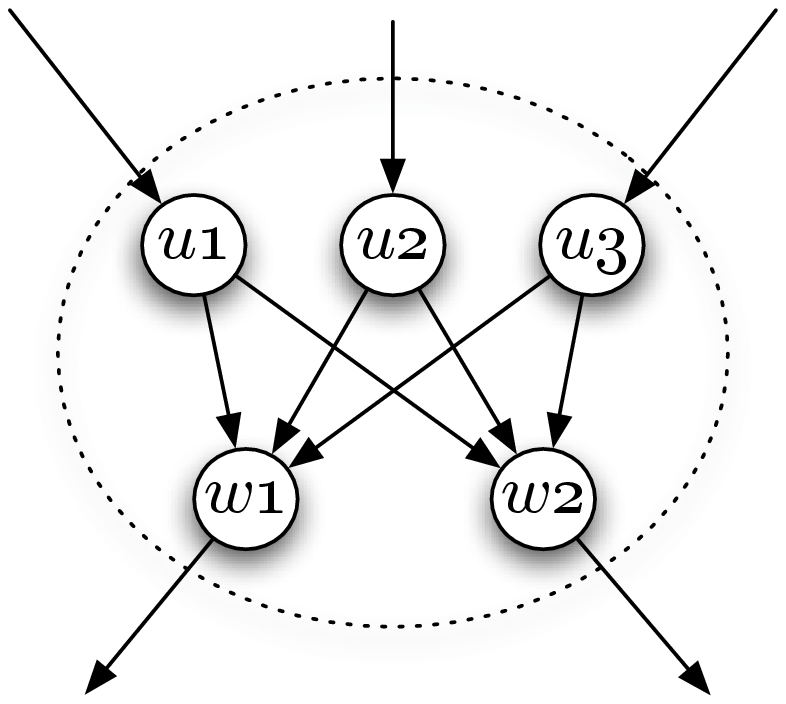}
\label{fig:decomp2}}
}
\caption{Decomposition of a node with $d_{in}=3$ and $d_{out}=2$.}
\label{fig:decomp}
\end{figure}

In the first stage of the algorithm ([C1] in Fig. \ref{fig:flowcentral}), we decompose each merging node that is not a sink as follows. (For a sink node with nonzero out-degree, introduce a virtual sink connected via $R$ links and decompose the original sink.) Consider a merging node $v$ with $d_{in}(\geq 2)$ incoming links and $d_{out}$ outgoing links (see Fig. \ref{fig:decomp}). We introduce $d_{in}$ nodes $u_1$, ..., $u_{d_{in}}$, which we call \emph{incoming auxiliary nodes}, and redirect the $i$-th ($1 \leq i \leq d_{in}$) incoming link of node $v$ to node $u_i$. Similarly, we create $d_{out}$ nodes $w_1$, ..., $w_{d_{out}}$, which we call \emph{outgoing auxiliary nodes}, and let the $j$-th ($1 \leq j \leq d_{out}$) outgoing link of node $v$ be the only outgoing link of node $w_j$. We then insert a link $(u_i, w_j)$ between each pair of nodes $u_i$ and $w_j$ ($1 \leq i \leq d_{in}$, $1 \leq j \leq d_{out}$). 

With each of the newly introduced links  between the auxiliary nodes, we associate a binary variable in a chromosome. Since each of those new links corresponds to a pair of connected incoming and outgoing links in the original graph, we can see a one-to-one correspondence between the binary variables here and those introduced in the algebraic method.

To verify the feasibility of a given chromosome $\underline{z}$, we first delete all the links associated with $0$ in $\underline{z}$ and then compute the max-flow between the source and each of the $d$ sinks. If an outgoing auxiliary node has only one incoming link, we can replace the incoming link, together with its outgoing link, by a single link without changing any of the max-flows. If an auxiliary outgoing node has no incoming link, we simply delete the node. If all $d$ max-flows are at least $R$, rate $R$ is achievable with network coding only at the outgoing auxiliary nodes with two or more incoming links. The number of remaining outgoing auxiliary nodes equals the number of coding \emph{links} in the original graph. Hence, by counting the number of such outgoing auxiliary nodes, we can calculate the fitness value $F$ of chromosome $\underline{z}$, defined as in \eqref{eq:fitnessDef}. 

The max-flow based evaluation of a single chromosome requires $O(d \cdot |E'|^2 \sqrt{|V'|})$ time \cite{AMO93}, where $|E'|$ and $|V'|$ are the number of links and nodes, respectively, in the decomposed graph. 

Note that, unlike the algebraic method, this feasibility test incurs no error. Since it works both with and without cycles, the graph decomposition method may be preferable to the algebraic method when centralized operation is feasible. Central operation requires that the topology of the whole network be known to a central computing node and may be slow. This approach may be appropriate, for example, in the planning stage of a network. However, the algebraic method plays a crucial role in the distributed version of the algorithm, as shown in Section \ref{sec:Distributed}.

\subsection{Computational Part of GA}\label{sec:CompPartGA}

In describing the computational part of the GA, we first discuss two novel elements designed specifically for the problem and then describe other typical GA components. The discussion applies for both the algebraic and graph decomposition methods.

\subsubsection{Block-Wise Representation and Operators}\label{sec:Block-wise}

A \emph{block} of length $k$ is defined to be a \emph{subset} of a chromosome consisting of $k$ binary variables that indicate the link states for the transmission onto a particular outgoing link fed by $k$ incoming links, i.e., the $k$ components of vector $a_j=(a_{ij})_{i\in\{1, ..., k\}}$ introduced in Section \ref{sec:Prob}. We may allow for a length-$k$ block to take all possible $2^k$ strings as in \cite{KAME06}, which we refer to as \emph{bit-wise representation}\footnote{In the GA community, the method for representing a candidate solution as a bit string is called \emph{genotype encoding}; we avoid the use of this term to minimize confusion with the term encoding in the context of network coding.}. Note, however, that once a block has at least two 1's, replacing all the remaining 0's with 1 has no effect on whether coding is done and that substituting 0 with 1, as opposed to substituting 1 with 0, does not hurt the feasibility. Therefore, for a feasible chromosome, any block with two or more 1's can be treated the same as the block with all 1's.

The above observation leads to \emph{block-wise representation}, where each block of length $k$ is allowed to take one of the following $(k+2)$ strings: [000...0] (1 string for \emph{no} transmission state), [100...0], [010...0], [001...0], ..., [000...1] ($k$ strings for \emph{uncoded} transmission state), [111...1] (1 string for \emph{coded} transmission state). If we let $w$ be the total number of blocks (i.e., $w=\sum_{v \in \mathcal{V}} d^v_{out}$) and $k_i$ denote the length of the $i$-th block $(i=1, ..., w)$, the search space size is reduced to $\prod_{i=1}^{w} (k_i+2)$, from $2^{\sum_{i=1}^w k_i}$ in the case of bit-wise representation. 

To preserve the structure of block-wise representation, we need a set of new genetic operators. In \cite{KAME06}, uniform crossover \cite{Mit96} and binary mutation \cite{Mit96} are used, which for comparison we refer to as \emph{bit-wise} operators\footnote{For uniform crossover, a pair of chromosomes exchanges each bit independently with a given probability, and for binary mutation, each bit of a chromosome is flipped independently with a given probability.}. Let us now define \emph{block-wise} operators as follows. For block-wise uniform crossover, we let each pair of chromosomes subject to crossover exchange each full block, rather than each individual bit, independently. For block-wise mutation, we let each block of length $k$ subject to mutation take another string chosen uniformly at random out of the other $(k+1)$ allowed strings.

It is interesting to note that the benefit of the smaller search space size in fact comes at the price of losing the information on the blocks with partially active link states that may serve as intermediate steps toward an uncoded transmission state. Also, whereas the \emph{average} number of bits flipped by block-wise mutation of a length-$k$ block using mutation rate $\alpha$ is $\frac{4k^2}{(k+1)(k+2)} \alpha$, which is smaller than that by the bit-wise mutation ($k \alpha$), the probability that 2 or more bits are flipped is often much larger for block-wise mutation; this may negatively affect the GA's ability to improve the solution through fine random changes. Hence, the overall effect of block-wise representation and operators on the algorithm's performance is not easy to predict theoretically. Section \ref{sec:Exp} includes an experimental evaluation of this question.

\subsubsection{Greedy Sweep}

We introduce another novel operator, referred to as \emph{greedy sweep}, where we inspect the best chromosome obtained at the end of the iteration and switch each of the remaining 1's to 0 if it can be done without violating feasibility ([C8] in Fig. \ref{fig:flowcentral}). This procedure can only improve the solution, and sometimes the improvement is substantial. Moreover, if we denote by $\underline{z}$ the chromosome after the greedy sweep, then $\underline{z}$ gives an upper bound on the number of coding links the same as in \cite{LSB06}.

\begin{lem}
The number of coding links associated with $\underline{z}$ is upper bounded by $R^3 d^2$ for an acyclic network and $(2B+1)R^3 d^2$ for a cyclic network, where $B$ is the minimum number of links that must be removed from the network in order to eliminate cycles.
\label{lem:Bound}
\end{lem}

\noindent \emph{Proof}:
Let us consider the graph decomposition method. Here, switching 1 to 0 in a chromosome implies that we delete the associated link in the decomposed graph. In the decomposed graph associated with $\underline{z}$, there is no link between auxiliary links that can be removed without violating the achievability. One can easily verify that there is a one-to-one correspondence between the links between auxiliary nodes in our decomposed graph (see Fig. \ref{fig:decomp}) and the set of the \emph{paths within} the gadget $\Gamma _v$ (see Fig. 2 in \cite{LSB06}). Now we can replace all non-merging nodes with a degree larger than 3 by the gadgets and greedily remove links in the same way as above, which, however, is irrelevant of the number of coding links. Therefore, from $\underline{z}$ we can construct a simple instance, as defined in \cite{LSB06}, and it gives the desired upper bounds on the number of coding links (Lemma 14 in \cite{LSB06}). \hfill $\square$

Lemma \ref{lem:Bound} provides our algorithm with a guarantee on its performance which is at least no worse than that of the algorithm in \cite{LSB06}.

\subsubsection{Typical GA Components}

When initializing the population ([C2] in Fig. \ref{fig:flowcentral}), we randomly generate each block and insert an all-one vector, whose effect as a feasible starting point is crucial as discussed in \cite{KAME06}. The iteration is terminated if the generation number reaches the predefined limit ([C4] in Fig. \ref{fig:flowcentral}). For selection ([C5] in Fig. \ref{fig:flowcentral}), we employ \emph{tournament selection} \cite{Mit96}, where we repeat a tournament between a predefined number of randomly selected chromosomes out of which the best one is selected (with replacement) for the next generation.

\section{Distributed Approach}\label{sec:Distributed}

Noting that the main advantage of network coding based multicast is that an efficient capacity-achieving code can be constructed in a distributed fashion \cite{HKM03, LMHK04}, a motivation for decentralization of the algorithm becomes apparent. That is, given that the actual multicast can proceed in a decentralized manner, an algorithm used for resource optimization should be more desirable if it does not require centralized computation.

Moreover, as will be discussed below, such decentralization enables the most time consuming task, fitness evaluation, to be distributed over the network such that the computational complexity required at each node depends only on local parameters. The size of the population often serves as an important factor for the ability of a GA to find a good solution \cite{HCG99}. Though it is not an easy task to predict the accurate population size required for a specific problem, it is always desirable to devise an evaluation method with a low complexity, which allows for a flexibility in adopting a large-sized population when needed.

\begin{figure}[h]
    \centering
    \includegraphics[height=1.8in]{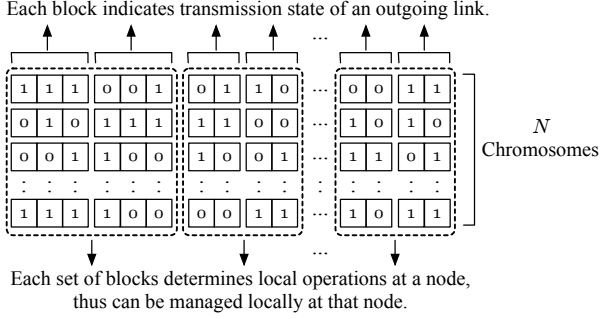}
    \vspace{-0.1in}
    \caption{Structure of Population}
    \label{fig:2}
\vspace{-0.1in}
\end{figure}

In this section, we present a novel distributed framework for our evolutionary algorithm, in which the feasibility test is done locally at each sink while the intermediate nodes actually construct random linear codes. With a limited amount of feedback information from the sinks and the merging nodes, fitness evaluation can be done with a substantially lower complexity. Furthermore, the population can be managed in a distributed manner such that each merging node locally manages a subset of the population that determines the local operations at that node (see Fig. \ref{fig:2}). Also, it will be shown that, with some amount of coordination, all genetic operations can be done locally at individual merging nodes. 

In addition to computational efficiency, this distributed approach has an important benefit that the coding resource optimization can be done on the fly while a network is operational, allowing for the following network coding protocol: As the source node sends an ``optimize" signal, all the nodes participating in the multicast go into the optimization mode, and as the distributed evolutionary algorithm proceeds, the links/nodes where coding is not required are identified. When the source node sends a ``transmit" signal, the network starts to multicast data based on the best network code found, in which coding is done only at the required links/nodes.

Since the distributed version of the algorithm is based on the algebraic method described in Section \ref{sec:Algebraic}, we need to take different approaches depending on whether the network has cycles or not. In the following, we begin to describe the details of the distributed approach assuming that the network is acyclic, and later in this section we extend the approach to cyclic networks, highlighting the changes to be made.

\subsection{Assumptions and Preliminaries}
\label{sec:AssumPrelim}

We assume that each link can transmit one packet of a fixed size per time unit in the given direction. Each link is also assumed to be able to send some amount of feedback data, typically much smaller than the packet size, in the \emph{reverse} direction. Also, we assume that each interior node operates in a burst-oriented mode; i.e., for the forward (backward) evaluation phase, each node starts updating its output only after an updated input has been received from all incoming (outgoing) links.

\begin{figure}[h]
    \centering
    \includegraphics[height=2in]{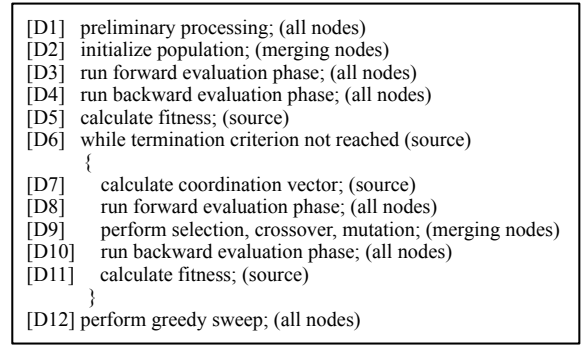}
    \vspace{-0.1in}  
    \caption{Flow of Distributed Algorithm}
    \label{fig:distributed}
\vspace{-0.1in}   
\end{figure}

The overall structure of our distributed algorithm is shown in Fig. \ref{fig:distributed} with the locations of each procedure specified. We now proceed to describe the detailed procedures of the algorithm in the order of their occurrences. 

\subsection{Details of Algorithm}

\subsubsection{\textbf{Preliminary Processing \emph{[D1]}}}

The source initiates the algorithm by transmitting the ``optimize" signal containing the following predetermined parameters: target multicast rate $R$, population size $N$, the size $q$ of the finite field to be used, crossover probability, and mutation rate. Each participating node that has received the signal passes the signal to its downstream nodes.

\subsubsection{\textbf{Population Initialization \emph{[D2]}}}

Let us consider a merging node with $d_{in} (\geq 2)$ incoming links. For \emph{each} of its $d_{out}$ outgoing links, the node has to manage a binary vector of length $d_{in}$, which we refer to as a \emph{coding vector}, to indicate the link states for a single chromosome. Hence, for the population of size $N$, the node must have $N \cdot d_{out}$ coding vectors to determine the operations at that node. To initialize this subset of the population, the merging node randomly generates $N \cdot d_{in} \cdot d_{out}$ binary numbers and set all the components to 1 for the coding vectors that correspond to the first of the $N$ chromosomes.

\subsubsection{\textbf{Forward Evaluation Phase \emph{[D3, D8]}}}

For the feasibility test of a chromosome, each node transmits a vector consisting of $R$ components, which we refer to as a \emph{pilot vector}. Each of its the components is from the finite field $\mathbb{F}_q$ and the $i$-th component represents the coefficient used to encode the $i$-th source data. We assume that a set of $N$ pilot vectors is transmitted together by a single packet.

The source initiates the forward evaluation phase by sending out on each of its outgoing links a set of $N$ random pilot vectors. Each non-merging node simply forwards all the pilot vectors received from its incoming link to all its outgoing links. 

Each merging node transmits on each of its outgoing links a random linear combination of the received pilot vectors, computed based on the node's coding vectors as follows. Let us consider a particular outgoing link and denote the associated $d_{in}$ coding vectors by $v_1$, $v_2$, ..., $v_{d_{in}}$. For the $i$-th ($1 \leq i \leq N$) output pilot vector $u_i$, we denote the $i$-th input pilot vectors received form the incoming links by $w_1$, $w_2$, ..., $w_{d_{in}}$. Define the set $J$ of indices as
\begin{align}
J = \{ 1 \leq j \leq d_{in} | \text{ the $i$-th component of $v_j$ is 1}\}.
\end{align}
Then,
\begin{align}
u_i = \sum_{j \in J} w_j \cdot \text{rand} (\mathbb{F}_q),
\end{align}
where rand$(\mathbb{F}_q)$ denotes a random element from $\mathbb{F}_q$. If the set $J$ is empty, $u_i$ is assumed to be zero.

\subsubsection{\textbf{Backward Evaluation Phase \emph{[D4, D10]}}}

To calculate a chromosome's fitness value, two kinds of information need to be gathered: 1) whether each sink can decode data of rate $R$ and 2) how many links are used for coding at each merging node. 

Each sink can determine whether data of rate $R$ is decodable for each of the $N$ chromosomes by computing the rank of the collection of received pilot vectors. It is worth to point out that this is the same algebraic evaluation method described in Section \ref{sec:Algebraic}, with the difference that, rather than computing the system matrix with randomized elements centrally, we now actually construct random linear codes over the network in a decentralized fashion. Hence, this feasibility test also bears the same, but uncritical, possibility of errors as in the centralized case. Regarding the number of coding links, each merging node can simply count the number links where coding is required by inspecting its coding vectors used in the forward evaluation phase.

For the feedback of this information, each node transmits a vector consisting of $N$ components, which is referred to as a \emph{fitness vector}. Each of the components must be at least $\lceil \log(|E|+2) \rceil$ bits long since for each chromosome the number of coding links can range from zero to $|E|$ and an additional symbol (infinity) is needed to signify infeasibility. The backward evaluation phase proceeds as follows:
\renewcommand{\labelenumi}{$\bullet$}
\begin{enumerate}
\item After the feasibility tests of the $N$ chromosomes are done, each sink generates a fitness vector whose $i$-th ($1 \leq i \leq N$) component is zero if the $i$-th chromosome is feasible at the sink, and infinity otherwise. Each sink then initiates the backward evaluation phase by transmitting its fitness vector to all of its parents.
\item Each interior node calculates its own fitness vector whose $i$-th ($1 \leq i \leq N$) component is the number of coding links at the node for the $i$-th chromosome plus the sum of all the $i$-th components of the received fitness vectors. Each node then transmits the calculated fitness vector to \emph{only one} of its parents, and an all-zero fitness vector (for just signaling) to the other parent nodes. 
\end{enumerate}
Note that, since the network is assumed to be acyclic, each coding link of a chromosome contributes exactly once to the corresponding component of the source node's fitness vector, and thus the above update procedure provides the source with the correct total number of coding links.
 
\subsubsection{\textbf{Fitness Calculation \emph{[D5, D11]}}}

The source calculates the fitness values of $N$ chromosomes simply by performing component-wise summation of the received fitness vectors. Note that if an infinity were generated by \emph{any} of the sinks, it should dominate the summations all the way up to the source, and thus the source can calculate the correct fitness value for the infeasible chromosome.

\subsubsection{\textbf{Termination Criterion \emph{[D6]}}}

The source can determine when to terminate the optimization by counting the number of generations iterated thus far. 

\subsubsection{\textbf{Coordination Vector Calculation \emph{[D7]}}}

Since the population is divided into subsets that are managed at the merging nodes, genetic operations also need to be done locally at the merging nodes. However, some amount of coordination is required for consistent genetic operations throughout all the merging nodes; more specifically, for 1) consistent selection of chromosomes, 2) consistent paring of chromosomes for crossover, and 3) consistent decision on whether each pair is subject to crossover. This information is carried by a \emph{coordination vector}, calculated at the source, consisting of the indices of selected chromosomes that are randomly paired and 1-bit data for each pair indicating whether the pair needs to be crossed over. The coordination vector is \emph{transmitted together with} the pilot vectors in the next forward evaluation phase.

\subsubsection{\textbf{Genetic Operations \emph{[D8]}}}

Based on the received coordination vector, each merging node can locally perform genetic operations and renew its portion of the population as follows:
\renewcommand{\labelenumi}{$\bullet$}
\begin{enumerate}
\item For selection, each node only retains the coding vectors that correspond to the indices of selected chromosomes. 

\item For block-wise crossover, each node independently determines whether each block is crossed over. Since no block is shared by multiple merging nodes, this can be done independently at each merging node.

\item For block-wise mutation, each node independently determines whether each block is mutated without any coordination with other nodes.
\end{enumerate}

\subsubsection{\textbf{Greedy Sweep \emph{[D12]}}}

Greedy sweep requires an additional protocol where the source is notified of the merging nodes with at least one coding link in the best solution obtained at the end of the iteration. Then, for each of such merging nodes, the source sends out a packet to test if uncoded transmission is possible on the link(s) where currently coding is required. Since this additional protocol requires more extensive coordination between nodes, we may leave this procedure optional, whose detailed description is omitted owing to space limitations. Note, however, that in our experiments the solutions obtained with the \emph{block-wise} representation/operations are already good enough so that further improvement by greedy sweep has never been observed. Nevertheless, greedy sweep may be useful as a safeguard that prevents the algorithm's poor performance due to misadjusted parameters, e.g., too small population size.

\subsection{Complexity}\label{sec:Complexity}

For evaluation of a single chromosome, each merging node $v$ computes random linear combinations of inputs in the forward evaluation phase, which requires $O(d^v_{in} \cdot d^v_{out} \cdot R)$, and each non-merging node $w$ simply forwards the received data, which requires $O(d^w_{out})$. Feasibility test at each sink $t$ is done by calculating the rank of a $d^t_{in} \times R$ matrix, where we assume $d^t_{in} \geq R$, hence it requires $O({d^t_{in}}^2 R)$. In the backward evaluation phase, update of a fitness vector takes $O(d^v_{in} + d^v_{out})$. Therefore, the computational complexity required for evaluation of a single chromosome is $O(\sum_{v \in \mathcal{V}} d^v_{in}d^v_{out}R + \sum_{w \in V \setminus \mathcal{V}} d^w_{out} + \sum_{t \in T} {d^t_{in}}^2 R)$, which can be substantially less than that for the centralized version of the algorithm.


\renewcommand{\theenumi}{\roman{enumi}}
\renewcommand{\labelenumi}{\theenumi)}

\subsection{Networks with Cycles}\label{sec:DistributedCyclic}

Cycles can be dealt with in two different ways as in other network coding problems. First, we can select a subgraph that does not contain a directed cycle, based on which we proceed to code construction and decoding in essentially the same manner as in the acyclic case \cite{JSC05}, \cite{HLKM05}. Alternatively, we may directly apply coding over cycles by combining information from possibly different time periods at intermediate nodes and deploying memory at the receivers for decoding \cite{KM03}, \cite{HMSE03}, \cite{EF05}, where the network code can be considered essentially a convolutional code.

The former of the above two scenarios allows for simple coding and decoding, but it may necessitate coding at the links/nodes where coding is not necessary if some link connections were not removed in the earlier stage \cite{KAME06}. On the other hand, the latter scenario may allow us to explore the full-fledged tradeoff between coding and capacity, but both specifying and decoding the code are more complex than in the former case.

Here we focus on the first scenario and describe how our distributed algorithm can be incorporated in the whole framework of such network coding schemes. Note that, if the original coding scheme is designed to operate on an acyclic subgraph selected beforehand, which seems more practicable, there is no reason to employ more complex network codes based on the original cyclic graph to minimize coding resources. However, we expect that a similar approach can be readily applied to the convolutional network coding scenario with an appropriate cycle-avoding mechanism for the transmission of the control messages such as the feedback information.

To set up an acyclic set of connections on a given network, we use the distributed algorithm in \cite{HLKM05}, where a binary variable is assigned to each pair $(l,l')$ of incident links indicating that the connection from link $l$ to link $l'$ is allowed or not. The value of each binary variable is determined such that the transmission along a directed cycle is prohibited. It is interesting to note that those binary variables used for subgraph selection are assigned to the link coefficients the same way as in our algorithm. Hence, our algorithm can be incorporated into the whole framework as follows:

\begin{enumerate}
\item Use the algorithm in \cite{HLKM05} to select the set of link coefficients to be used for transmission.
\item Each node then exchanges the binary variables assigned to its links with its neighbors so that each node can identify the allowed connections.
\item We then apply the our distributed algorithm ignoring the link coefficients that correspond to the disallowed connections.
\end{enumerate}

Alternatively, if minimizing the link cost is our primary concern, we may use the algorithms that find the minimum cost subgraph in a decentralized fashion, such as the one in \cite{LMHK04}. The resulting subgraph does not contain a directed cycle if the link costs are all positive. Hence, a two-stage method is possible where the minimum cost subgraph selection is followed by our distributed algorithm. This two-stage method may perform very well in practice, as will be demonstrated in the next section.
 
\section{Experimental Results}\label{sec:Exp}

The parameters used for the experiments are as follows: Population size is 150 and the iteration terminates after 1000 generations. Tournament size (for selection) and mutation rate are 100 and 0.012, respectively, for the block-wise case, and 10 and 0.006 for the bit-wise case. Crossover rate is fixed at 0.8.

\subsection{Effects of Block-Wise Representation and Operators}

We evaluate the performance of our algorithm with the block-wise and the bit-wise representations and operators, using the centralized version of the algorithm with the graph decomposition method. The experiments are based on the two topologies generated by the algorithm in \cite{MP04} with the following parameters: (50 nodes, 87 links, 10 sinks, rate 5) and (75 nodes, 156 links, 15 sinks, rate 7).

For comparison, we also perform experiments with two existing greedy approaches by Fragouli et al. \cite{FS06} ("Minimal 1") and Langberg et al. \cite{LSB06} ("Minimal 2"). For both approaches, link removal is done in a random order. For Minimal 1, the subgraph is selected also in a greedy fashion by sequentially removing links. Table \ref{tab:Eval1} shows the best and the average values, as well as standard variation, obtained in 30 trials.

\begin{table}[h]
\centering{
\begin{tabular}{|c|c|c|c|c|c|c|} \hline
    & \multicolumn{3}{c|}{(50,87,10,5)} & \multicolumn{3}{c|}{(75,156,15,7)}
    \\ \cline{2-4} \cline{5-7}
\raisebox{1.5ex}[0cm][0cm]{}
    & Best  & Avg. & Std. & Best  & Avg. & Std.\\ \hline
Block-wise
    & 2     & 2.40 & 0.62  & 3     & 3.63 & 0.61\\ \hline
Bit-wise
    & 2     & 3.33 & 1.03  & 5     & 6.43 &1.30\\ \hline
Minimal 1
    & 3     & 4.90 & 1.37  & 6     & 9.50 & 2.16\\ \hline
Minimal 2
    & 3     & 4.33 & 1.37  & 4     & 7.90 & 1.71\\ \hline
\end{tabular}
}
\caption{Number of Required Coding Links}
\label{tab:Eval1}
\vspace{-0.2in}   
\end{table}

Tabel \ref{tab:Eval1} shows that the block-wise representation and operators clearly outperform the bit-wise counterpart in all aspects. We can also observe that the performance of our algorithm, with either of the two representation and operators, is at least as good and often better than that of both Minimal 1 and Minimal 2, except only in the best value of the bit-wise case for the larger network. More in-depth comparisons between these algorithms can be found in our subsequent paper \cite{KAOM07}, where our algorithm with the block-wise representation and operators is found to exhibit a far greater performance advantage over the other three cases.

\subsection{Performance of Distributed Algorithm}

Since the distributed algorithm shares the same computational part of GA with the centralized one, the two algorithms show the same performance in terms of solution quality. However, as shown in Section \ref{sec:Complexity}, the computational complexity required by the distributed algorithm depends on local topological parameters, and this can often lead to a significant gain in terms of the running time.

To compare the running times of the two approaches, we generate a set of highly connected topologies such that there exists a link between each pair of nodes $i$ and $j$ ($i<j$), where the source is node 1 and the sinks are the last 10 nodes. This test is pessimistic in the sense that the distributed algorithm is simulated on a single machine while each node's function is performed by a separate thread, thus it cannot benefit from the multi-processing gain whereas it only suffers from additional computational burdens for managing a number of threads. Table \ref{tab:Eval3} shows that, nevertheless, the distributed algorithm exhibits an advantage in running time as the size of the network grows.

\begin{table}[h]
\centering{
\begin{tabular}{|c|c|c|c|c|c|c|} \hline
    Number of nodes & 15 & 20 & 25 & 30 & 35 & 40 \\ \hline
    Centralized & 0.3 & 1.5 & 4.3 & 13.5 & 29.5 & 65.6 \\ \hline
    Distributed & 1.8 & 2.7 & 4.4 & 6.3 & 10.8 & 15.4 \\ \hline
\end{tabular}
}
\caption{Elapsed Time Per Generation (seconds)}
\label{tab:Eval3}
\vspace{-0.2in}   
\end{table}

\subsection{Effectiveness of Two-Stage Method}\label{sec:TwoStage}

We introduced in Section \ref{sec:DistributedCyclic} a two-stage method where we first select a minimum-cost subgraph assuming network coding is done everywhere and then apply our evolutionary algorithm to the resulting subgraph. Though not optimal, this two-stage method can be very useful when optimization over both link cost and coding cost is required; the minimum link cost is guaranteed, and the resulting acyclic subgraph can often be substantially smaller than the original network.

We test the two-stage method on ISP 1755 and 3967 topologies from the Rocketfuel project \cite{SMW02}, using the algorithm in \cite{LMHK04} to obtain a minimum cost subgraph. With 10 randomly selected sinks and target rates 2, 3, and 4 on both topologies, each of 30 runs always ends up with \emph{zero} coding links. These results may suggest that, while \emph{assuming} network coding enables to calculate a minimum-cost subgraph, there may be very few links/nodes where network coding is actually required in the end.

\section{Conclusions and Future Work}\label{sec:Con}

We have presented evolutionary approaches to minimizing the resources used for network coding in a single multicast scenario. The proposed algorithms have been shown to have advantages over our previously proposed algorithm \cite{KAME06}, as well as other existing greedy algorithms, in terms of the applicability to general topologies, the solution quality, and the practicability in a distributed environment. 

For future research, we may further improve the distributed algorithm, by a smarter management of population and packet transmissions, such that it converges faster to a better solution and works asynchronously, providing robustness against delay, failure, or topological changes in the network. Also, we could observe a tradeoff between coding and link usage in the sense that in some networks \cite{KAME06}, reducing link usage first by subgraph selection may increase coding in the remaining subgraph, which is not the case in our experiments in Section \ref{sec:TwoStage}. Hence, whether there exists such a tradeoff can be considered a topological property of a network and thus the effectiveness of the two-stage method discussed in Section \ref{sec:DistributedCyclic} may depend on the network topology. We may further investigate this tradeoff using evolutionary algorithms designed for multi-objective optimization.


\section*{Acknowledgment}
The first author would like to thank Fang Zhao for her help on the experiments.



\bibliographystyle{IEEEtran}
\bibliography{IEEEabrv,mrabbrev,Infocom07}
%

\end{document}